# Zero-energy band observation in an interfacial chalcogen-organic network

Yichen Jin[1], Ignacio Gonzalez Oliva[1], Hibiki Orio[2], Guangyao Miao[1,3], Maximilian Ünzelmann[2], José D. Cojal González[1], Angelina Jocic[4], Yan Wang[3], Xiaoxi Zhang[3], Jürgen P. Rabe[1], Kai Rossnagel[5,6], Milan Kivala[4], Claudia Draxl[1], Friedrich Reinert[2], Carlos-Andres Palma[1,3]

*1 Department of Physics and CSMB, Humboldt-Universität zu Berlin, 12489 Berlin, Germany*

*2 Experimentelle Physik VII and Würzburg-Dresden Cluster of Excellence ct.qmat, Universität Würzburg, Am Hubland, D-97074 Würzburg, Germany*

*3 Institute of Physics, Chinese Academy of Sciences, 100190 Beijing, China*

*4 Organisch-Chemisches Institut, Universität Heidelberg, 69120 Heidelberg, Germany*

*5 Institute of Experimental and Applied Physics, Kiel University, 24098 Kiel, Germany*

*6 Ruprecht Haensel Laboratory, Deutsches Elektronen-Synchrotron DESY, D-22607 Hamburg, Germany*

**Abstract:** Structurally-defined molecule-based lattices such as covalent organic or metal-organic networks on substrates, have emerged as highly tunable, modular platforms for two-dimensional band structure engineering. The ability to grow molecule-based lattices on diverse platforms, such as metal dichalcogenides, would further enable band structure tuning and alignment to the Fermi level, which is crucial for the exploration and design of quantum matter. In this work, we study the emergence of a zero-energy band in a triarylamine-based network on semiconducting 1T-TiSe$_2$ at low temperatures, by means of scanning probe microscopy and photoemission spectroscopy, together with density-functional theory. Hybridization between the position-selective nitrogens and selenium $p$-states results in CN–Se interfacial coordination motifs, leading to a hybrid molecule-semiconductor band at the Fermi level. Our findings introduce chalcogen-organic networks and showcase an approach for the engineering of organic-inorganic quantum matter.

# Introduction

On-surface molecule-based fabrication is one of the most versatile methods for the design of extended carbon nanomaterials (*1-4*). In particular, molecule-based networks fabricated through covalent organic, metal-organic, and hydrogen-bonded bonding schemes(*5-8*) have led to the engineering of topological states (*9-12*), such as Dirac bands (*13, 14*) and flat bands (*15-18*). Band structure characterization thus far has been reported mostly on metal substrates, often hindering accessibility to the molecular states, particularly near the Fermi level (*19-23*). To overcome this limitation, several strategies, such as the intercalation of insulating layers (*8*) or the stacking of molecular multilayers (*24-26*), have been used to study electronic bands of molecular origin (molecular bands). A doping or pinning approach is also commonly employed to tune the molecular bands around the Fermi level(*27*). However, this typically results in non-zero injection barriers with respect to the metallic Fermi level and overlapping with surface states (*21, 28-31*). Advances in the preparation of van der Waals (vdW) materials (*32-34*) are expected to expand the scope of on-surface molecular fabrication, leading to adsorbate states (*35, 36*) and the emergence of new interlayer states (*37-39*). This is a crucial step toward experimentally resolving the properties of novel molecular bands and enabling the use of molecule-based fabrication technologies based on the vdW semiconductors and self-assembly. For example, it has been reported for vdW interfaces between molecules and conductive substrates, the preservation of magnetic (*40, 41*), vibrational, (*42*) and electronic (*8*) properties, as well, as an overall reduction of the substrate-molecule interactions (*43, 44*).

Altogether, molecular bands at zero-energy have yet to be observed experimentally on insulating or semiconducting vdW substrates, which is a fundamental step towards fabricating novel quantum materials, such as molecule-based Mott insulators, metals, half-filled spin-lattices and corresponding electronic devices (*45, 46*). Here we introduce an interfacial chalcogen-organic network, specifically a vdW hybrid heterostructure consisting of planarized triarylamine network (*47, 48*) decorated with three cyano (CN) functions, combined with the transition metal dichalcogenide (TMDC) 1T-$TiSe_2$, which we refer to as triarylamine/$TiSe_2$ (Scheme 1) or simply Tria/$TiSe_2$. The structural characteristics of the triarylamine/$TiSe_2$ interface were resolved through low-temperature scanning tunneling microscopy (STM) at 5.1 K. The electronic properties are investigated by joint scanning tunneling spectroscopy (STS), and angle-resolved photoelectron spectroscopy (ARPES) measurements. The experimental results are then interpreted by density-functional theory (DFT). This combined approach using STS, ARPES and DFT identifies a zero-energy band stemming from the hybridization between the molecules and the TMDC. Our results introduce a foundational strategy for engineering zero-energy molecular bands in heterostructures.

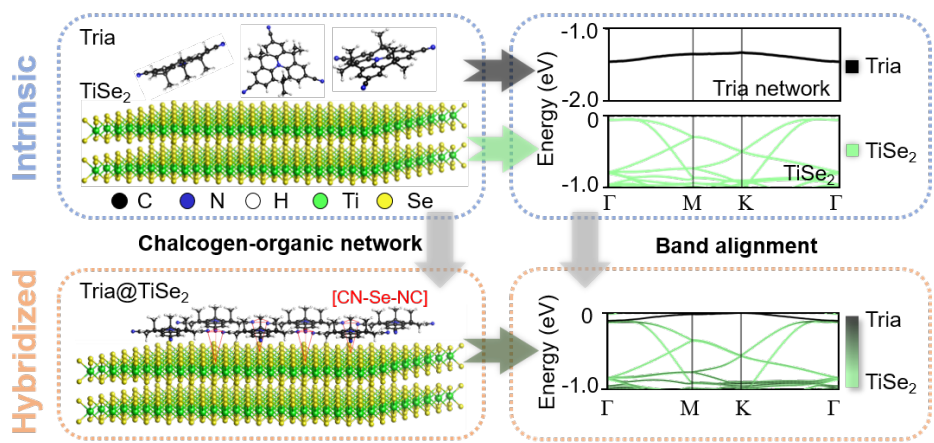

**Scheme 1**: Triarylamine self-assembly on TiSe$_2$ forming a chalcogen-organic network through a CN–Se and N–Se interfacial coordination motifs (left panel). The alignment of the valence band of triarylamine (Tria, *4,4,8,8,12,12-hexamethyl-4H,8H,12H-benzo[1,9]quinolizino[3,4,5,6,7-defg]acridine-2,6,10-tricarbonitrile*) and TiSe$_2$ results in a molecular band at the Fermi level (right panel).

# Results

**Lattice and spectroscopy:** The structural parameters of triarylamine/TiSe$_2$ are studied by high-resolution STM. The deposition (see Methods) of triarylamine at room temperature on the surface of TiSe$_2$ results in an ordered domain of the molecular structure as shown in Fig. 1A. In the high-resolution STM topography data, a single triarylamine features three light and three dark triangular motifs with an angular difference of 60°. Zoomed-in STM data and the corresponding apparent height profiles of the triarylamine and TiSe$_2$ regions are presented in Figs. 1B to 1E. The brighter triangular motif is assigned to the dimethylmethylene bridges of triarylamine, while the CN groups at the edges of the triangles show lower contrast and a more diffuse distribution (see Fig. S1). These features are slightly different from the results of triarylamine on Au(111) (*47-49*), where the CN groups are rarely visible.

STM data reveal a lattice constant of $a_{triarylamine}$ = 1.460 ± 0.003 nm and $a_{TiSe2}$ = 0.350 ± 0.002 nm, with an interlayer distance of 0.28 nm (Fig. S2). The primary repeat unit consists of a 4 × 4 supercell of TiSe$_2$ and a unit cell of triarylamine, resulting in a small lattice mismatch of 4 ± 1 %. The incomplete lattice commensurability is evident from the slight rotation of approximately 20% of the molecules away from the three-fold symmetry with their neighbors. In Fig. 1B, a symmetric trimeric motif arising at triple molecular junctions by CN groups emerges among three adjacent molecules. This arrangement differs from an H-bonded propeller-like motif [CN···H] observed in other triarylamine-based interfaces (*48, 50*). Notably, the TiSe$_2$ lattice at low temperature undergoes a reconstruction into a 2×2 charge density wave (CDW) (*51*) (Fig. 1C), which could modulate the apparent disorder of non-trimeric motifs.

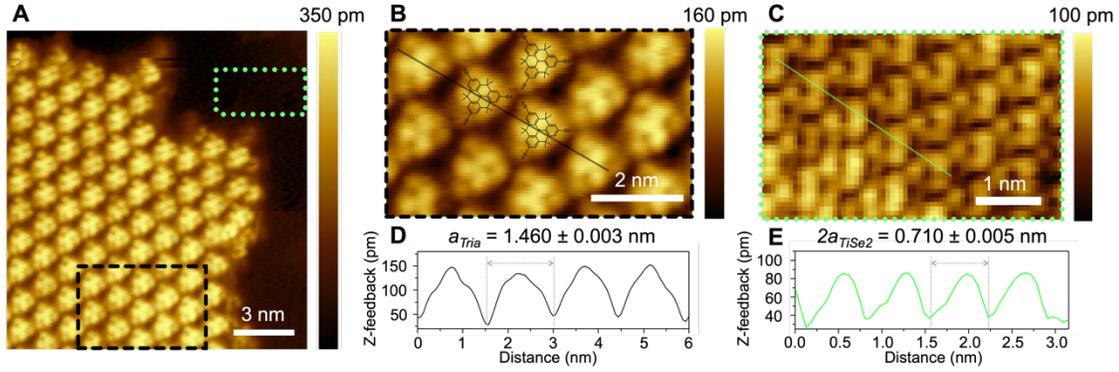

**Figure 1: Topography of triarylamine on TiSe$_2$.** **(A)** High-resolution STM image of submonolayer crystallized triarylamine on TiSe$_2$. The black and green squares represent the zoomed-in area. (U$_{sample\_bias}$ = 100 mV, set-point = 2 pA, and T = 5.1 K) **(B, C)** Zoom into areas from (A) demonstrating the crystallized triarylamine and the CDW in the TiSe$_2$ substrate. **(D, E)** Height profiles **corresponding** to the black and green lines in B and C, respectively.

The spectroscopic characterization of the electronic properties was first performed with ARPES measurements at 15 K (Fig. 2A). At a binding energy (E$_b$) of approximately -0.5 eV, a spectral signature assigned to a folded Se 4$p$ band is observed for both pristine TiSe$_2$ and Tria/TiSe$_2$. The lower intensity of this feature in the heterostructure is explained by the additional scattering effects introduced by the molecular layer (Fig. 2B). Another spectral signature is observed at the Fermi level, assigned to the Ti 3$d$ band. In the hybrid system, the enhancement of this signature suggests a small amount of charge transfer from triarylamine to TiSe$_2$. However, this alone does not directly indicate the presence of a molecular state at the Fermi level and could be influenced by interfacial or scattering effects. A new diffuse feature assigned to the molecular layer appears below E$_b$ ~ –1.1 eV, as observed in the Tria/TiSe$_2$ ARPES spectrum. The feature is expressed as a broadened shoulder in the energy distribution curve (EDC in Fig. 2A) integrated from $K_x$ = 0 (dotted line). Note that, the Se 4$p$ back-folded band appears at a deeper E$_b$ position due to the slight misalignment of the E-$K_x$ cut with the M point during the measurements ($K$-space position in Fig. 2C).

To further investigate the electronic properties of our system, STS measurements were performed at five different positions (labelled #1-5 in Fig. 2D) following two paths. The STS spectra for the five positions are shown in Fig. 2E. For positions #1-3, a localized zero-energy peak (FWHM~50 mV) is observed, while for position #5, the STS spectrum is from intrinsic TiSe$_2$ (*52*). As the tip moves away from the molecular layer, the intensity of the zero-energy peak gradually decreases, suggesting a bulk molecular origin. (For additional STS spectra, see Fig. S3.) The presence of a molecular state at the Fermi level is further clarified by first-principles calculations below, and is addressed in the Discussion section. Note that the weak conductivity signal in Fig. 2E can be attributed to the low tunneling set point (5 pA) on the gapped semimetal, leading to valence band conductivities much lower than that usually reported when employing set points of >100 pA in similar kind of materials on metal surfaces(*53-55*).

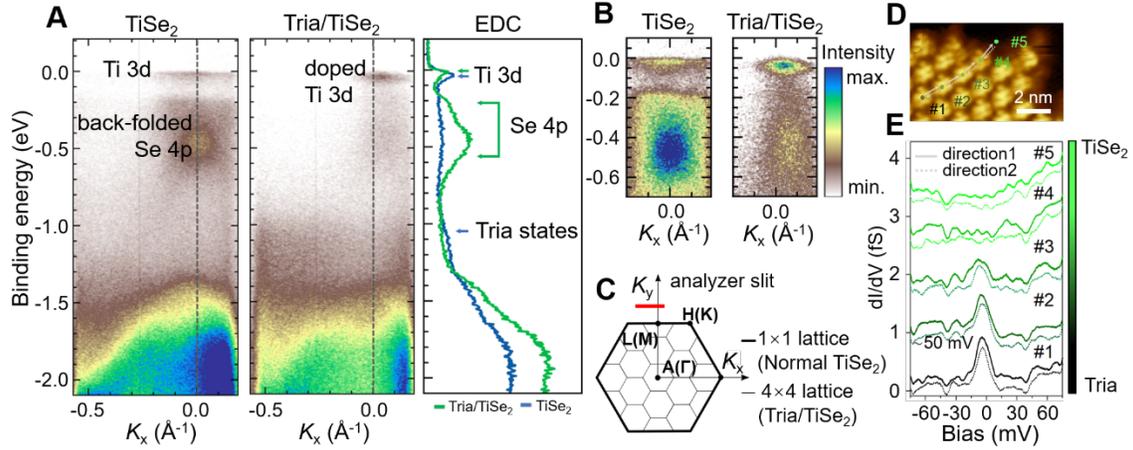

**Figure 2: ARPES and STS measurements to characterize the electronic structure of triarylamine on TiSe₂.** **(A)** ARPES intensity maps around the M point with He-I: 21.2 eV at 15 K: intrinsic TiSe$_2$, ~0.5 submonolayer triarylamine on TiSe$_2$ with the same contrast. The EDC is obtained near $K_x$ = 0.0 (dotted line). **(B)** Zoomed-in area of Ti 3$d$ states near the Fermi level in ARPES maps. **(C)** Illustration of the sample's *K*-space position during measurements. **(D)** Zoomed-in area with marks (from the large STM image in Fig. 1A) showing the STS measurement positions. **(E)** STS data on the center of triarylamine in the heterostructure at the indicated positions with two different reproducible measurement directions (set-point: 5 pA, $U_{sample\_bias}$ = –300 mV, $U_{ac}$ = 2 mV, f = 973 Hz and T = 5.1 K).

**Molecular band structure:** DFT calculations were conducted to investigate the structural and electronic properties of the heterostructure. Figure 3A shows the geometry used for the calculations, where a single molecule of triarylamine is adsorbed in a 4 × 4 × 1 TiSe$_2$ supercell structure. The calculated adsorption distance between triarylamine and TiSe$_2$ of 0.32 nm is in good agreement with the 0.28 nm from STM data (Fig. S2), indicating physical adsorption (*56*).

Figure 3B shows the band structure of triarylamine/TiSe$_2$ calculated with PBE. The adsorption of triarylamine manifests itself through the emergence of a band at around -0.1 eV in the valence region. Interestingly, the band is shifted to the Fermi level, as seen at the M and K points. The KS orbital at the Γ point (red dot in Fig. 3B) is visualized in Fig. 3C, revealing a localized molecular state and confirming that this level corresponds to the highest occupied molecular orbital (HOMO) of triarylamine. The top and side views of the KS orbital show that the isosurface extends over the entire molecule in real space with contributions from Ti and Se atoms. The latter highlights a hybridized character at the Γ point, which is not present for the HOMO at the M and K points (Fig. S4). Several molecular bands emerge within the energy window of –1.3 to –1.6 eV, which are assigned to the lower occupied molecular orbitals (LOMOs). Like in similar physisorbed systems with molecules on a TMDC (*57-59*), the KS band structure of the heterostructure resembles a superposition of the band dispersion of the constituents (Fig. S4).

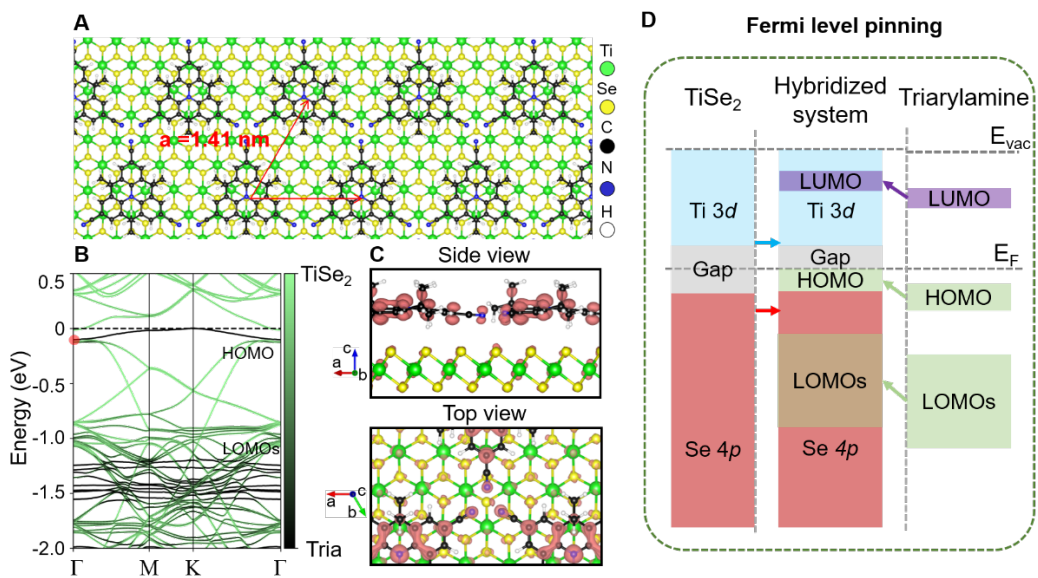

**Figure 3: DFT calculations of the electronic structures of triarylamine on TiSe$_2$.** (**A**) Structural model based on the STM images. (**B**) KS band structure of triarylamine on TiSe$_2$. The TiSe$_2$ (green) and triarylamine (black) states are projected onto the bands with the weights illustrating the contribution of the individual states. (**C**) Side and top view of the projected KS wave functions (red) of the triarylamine (HOMO) at the Γ point detailing the CN–Se (3.4 Å) interfacial coordination motif with an isosurface of $0.001e·Å^{-3}$. (**D**) Fermi level pinning model (*60*) of the hybrid system.

To further investigate the alignment of the HOMO to the Fermi level, the charge density difference, defined as $\Delta\rho = \rho_{hybrid} - (\rho_{substrate} + \rho_{molecule})$, quantifies the interfacial charge redistribution in the heterostructure. Figure 4A shows a localized charge transfer from the CN cluster at the triarylamine edge to the Se atom through its local [CN-Se] structure (N-Se: 3.4 Å leading to a short CN-NC interaction distance: 2.4 Å), such attraction between selenium and nitrogen has been recently observed in organoselenium complexes on surfaces (*61*). Integrating the highlighted region of the plane-averaged charge density rearrangement $\Delta\rho$ in Fig. 4B results in a charge transfer of –0.101 $e^-$ per molecule. Combined with the electron distribution in the KS orbitals (Fig. 3C), there is evidence of electronic redistribution on the Se atom. This suggests the presence of position-selective hybridization. The projected density of states (PDOS) in Figs. 4C and 4D also reveal subtle information about the interaction between molecular and substrate states in the valence band. Figure 4C shows that the atoms of both the molecule and TiSe$_2$ contribute to the

PDOS at the Fermi level. Resolving the PDOS per individual atom (Fig. 4D) allows us to identify that the HOMO has mainly contributions from the central nitrogen (N-2) and CN atoms, while the LOMOs have mostly contributions from the CN (N-1, N-3, N-4) and minimal to none from N-2.

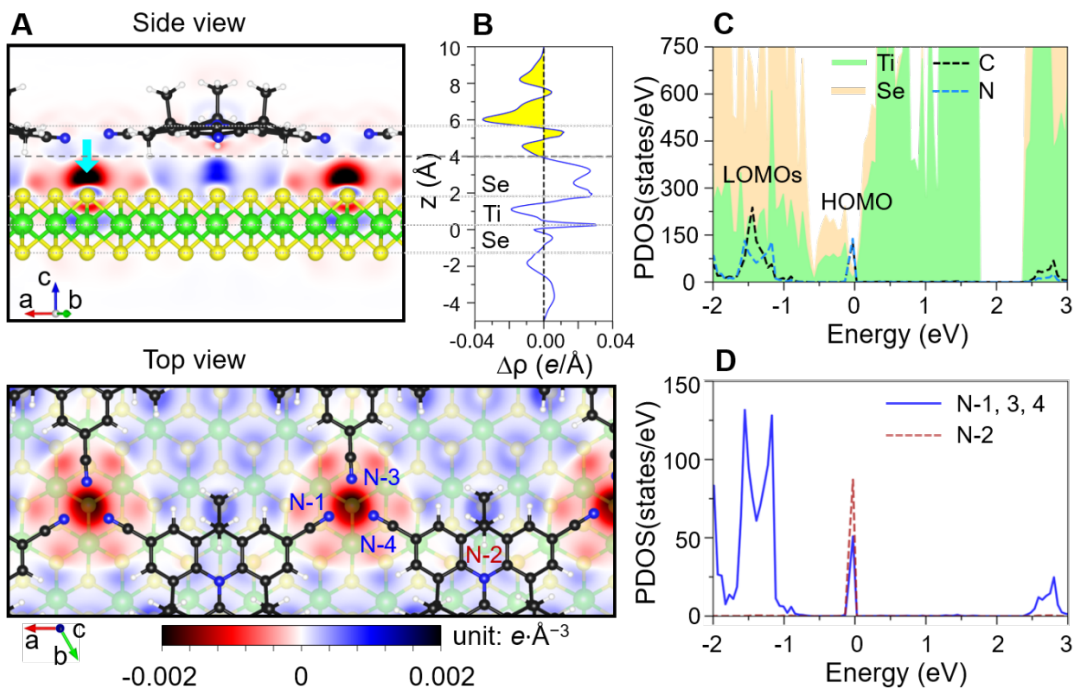

Figure 4: Electron distribution for the CN-Se interfacial coordination motifs. (A) Top and side view of the charge density difference of triarylamine on TiSe$_2$. The color code indicates negative (red) and positive (blue) values at an isosurface of $\pm 0.002\ e\cdot\text{Å}^{-3}$. The cyan arrow marks the direction of electron transfer. (B) Plane-averaged charge rearrangement $\Delta\rho$ due to the adsorption of triarylamine. A positive (negative) value of $\Delta\rho$ represents accumulation (depletion) of charge density. Integrating over the filled yellow areas results in the transferred molecular charge. (C, D) PDOS in states per eV resolved per species and per individual N atom (marked in A, top view) of triarylamine on TiSe$_2$.

## Discussion

As described in the ARPES and STS experiments, the data at zero-energy, i.e., Fermi level, are clearly influenced by the molecular layer. Spatially resolved STS measurements resolve a zero-energy peak (FWHM $\approx$ 50 mV) localized on the triarylamine (Fig. 2E). Similar molecular states with an FWHM of 61±6 mV have been reported near the Fermi level on graphene (*62*). Another example is the case of phthalocyanine on MoS$_2$ (*42*), where a narrow FWHM of approximately 30 mV for the zero-energy peak has been attributed to a protective effect of the TMDC gap on the state lifetime. The DFT band structure of

triarylamine/TiSe$_2$ shows that the HOMO appears close to the Fermi level (Fig. 3B), which supports the observation of a zero-energy peak in STS measurements. The position of the HOMO at Fermi is further supported by TiSe$_2$ high work function (5.7 eV) (*63*) aligning with occupied molecular states. In ARPES, the HOMO alignment at Fermi is only indirectly substantiated by diffuse features starting at an energy of –1.1 eV (Fig. 2A): DFT calculations reveal several molecular states at a similar energy window, which we refer to LOMOs for simplicity. The dense and flat LOMOs states are consistent with the weakly diffuse character in ARPES. The EDC shows a clear separation between the LOMOs and TiSe$_2$ states. The latter energies are likewise close to the calculated DFT values near the M point. The signal near the Fermi level is remarkably unresolved in ARPES and predominantly explained by an enhancement of the Ti 3*d* band due to charge transfer. This observation is also consistent with HOMO hybridization with the Se and Ti atoms, as shown by the KS orbital plot at Γ point (Fig. 3C). Moreover a limited contribution of the HOMO in the ARPES measurements is expected when comparing to the calculated PDOS contributions (cf. Fig. 4C).

DFT calculations reveal that a charge transfer channel occurs at periodic sites, primarily between the symmetric trimeric CN motifs and the Se atoms. This charge redistribution affects the in-plane self-assembly by favoring close CN–NC intermolecular distances. The central nitrogen adsorbs on the 1T hollow site and is thus farthest from the Se atoms. The CN–Se hybridization is therefore expected to modulate the band dispersion in this class of quantum materials. The charge transfer (–0.101 $e^-$ per molecule) may be further optimized by molecular design (e.g., regulating the number, geometry and substrate distance of cyano groups) or by substrate engineering (e.g., Se vacancy doping), providing a new platform for tailoring the spin, electronic and structural properties of molecule-based quantum matter.

In summary, the structural and electronic properties of interfacial chalcogen-organic networks involving cyano-functionalized triarylamine on 1T-TiSe$_2$ depend on two key parameters: the symmetric trimeric motif formed by their CN groups that control the self-assembly, and the strength of CN-chalcogen and N-chalcogen interaction which determines the electronic band properties. Modulating these two factors is a critical step toward designing hybrid systems with programmable quantum properties.

# Materials and Methods

**Sample preparation:** The 1T-TiSe$_2$ substrates were synthesized by the chemical vapor transport method with details to be found in (*64*). The cyano-substituted triarylamine was synthesized according to the previously reported procedure (*48*). Time-of-flight mass spectrometry (TOF-MS) confirmed the molecular purity, as depicted in Fig. S5. Before every measurement, the TiSe$_2$ substrates were cleaved in UHV condition (1 × 10$^{-8}$ mbar), followed by a quick flash at 473 K to remove the contaminants. For the on-surface synthesis of submonolayers, the triarylamine powder was deposited at a distance of ~15 cm at 413 K for 21 min.

**Measurements**: Low temperature STM experiments (4 K STM, *CreaTec GmbH*) were carried out at ~5 K under UHV conditions (2 × 10$^{-10}$ mbar). STM data were acquired with a Pt/Ir tip, and all STM images shown in this work were processed by the *Gwyddion* software (*65*). STS was measured with a lock-in amplifier (f = 973 Hz). TOF-MS (*Kore Technology Ltd*) spectra were calibrated based on a fitted line obtained from the peaks corresponding to $H_2O$ and $Si_3O_3C_5H_{15}$ under UHV conditions ($1 \times 10^{-9}$ mbar). ARPES and XPS measurements were performed in UHV chamber ($3 \times 10^{-10}$ mbar) using a hemispherical analyzer (*SIENTA omicron R3000*). ARPES spectra were conducted using non-monochromatized He-I (21.22 eV) light. XPS spectra were recorded with a non-monochromatized Mg-Kα (1253.6 eV) source. The photoelectron emission angle of XPS was 60° for surface sensitivity. XPS results support the hybridized interface and are discussed in the supplementary information in Fig. S6.

**DFT calculations**: To meet the structural conditions derived from the STM experiments as close as possible, we constructed a 4×4×1 supercell of pristine $TiSe_2$ monolayer, with an in-plane lattice parameter of 0.35 nm and a vacuum of 1.5 nm. We optimize the atomic structure, consisting of 106 atoms, using the all-electron code FHI-aims (*66*) until the interatomic forces were below a threshold value of 0.001 eV/Å$^3$ (*67, 68*). The subsequent ground-state calculations were performed using `exciting` (*69*), an all-electron full-potential code, implementing the family of linearized augmented plane wave plus local orbitals (LAPW+LO) methods. The generalized gradient approximation in the Perdew-Burke-Ernzerhof (PBE) parametrization (*70*) was used for exchange and correlation effects. The sampling of the BZ was carried out with a homogeneous 6×6×1 Monkhorst-Pack k-point grid, and the occupations determined by the extended linear tetrahedron method. To account for vdW forces between the substrate and molecules and for intermolecular interactions, we adopted the Tkatchencko-Scheffler (TS) method (*71*). Dipole corrections were also considered. The muffin-tin (MT) spheres in the inorganic component (Ti, Se) were chosen to have equal radii of 2.2 bohr. For the molecules, the radii are 0.9 bohr for hydrogen (H), 1.1 bohr for carbon (C), and 1.2 bohr for nitrogen (N). We used a basis-set cutoff of $G_{max} = 3.889$ bohr$^{-1}$. To reduce the computational cost, spin-orbit coupling (SOC) is not considered in this work. All the input and output files can be downloaded from NOMAD (*72*), URL: https://dx.doi.org/10.17172/NOMAD/2024.09.06-1.

# Acknowledgments

The computing time on the supercomputers Lise and Emmy at NHR@ZIB and NHR@Göttingen is gratefully acknowledged. We acknowledge the support at MAX IV from Yuriy Dedkov for the PES measurements. We thank Sebastian Tillack for calculation discussions and Meike Stöhr for critical feedback, material and organizational support for the project.
**Funding:**
Chinese Academy of Sciences QYZDBSSW-SLH038, XDB33000000, XDB33030300 (C.A.P.)
Beijing Natural Science Foundation (International Scientists Project) IS24031 (C.A.P.)
National Science Foundation of China 12474178 and 12304238 (C.-A.P., Y. J. and X.Z.)
DFG Cluster of Excellence 'Matters of Activity, Image Space Material' under Germany's Excellence Strategy EXC 2025 (J.C.G, C.-A.P. and J.P.R.)

**Author contributions:**
Ideas: Y.J., G.M., I.G.O. and C.-A.P.
STM experiment at HU: C.-A.P., J.C.G., Y.J. and J.P.R.
DFT calculations: I.G.O. and C.D.
PES experiments at JMU: H.O., M.Ü. and F.R.
TOS-MS measurements at IOP-CAS: Y.W. and X.Z.
Triarylamine molecule preparation: A.J. and M.K.
TiSe$_2$ crystal preparation: K.R.
Data analysis: Y.J., I.G.O., H.O., C.-A.P. and J.C.G. with the help of G.M.
Writing-original draft: Y.J. and I.G.O. with input from H.O.
Writing-review and editing: all authors.

**Competing interests:** Authors declare that they have no competing interests.

**Data and materials availability:** All data are available in the main text or the supplementary materials.

# References

1. J. V. Barth, G. Costantini, K. Kern, Engineering atomic and molecular nanostructures at surfaces. *Nature* **437**, 671-679 (2005).
2. A. Harada, R. Kobayashi, Y. Takashima, A. Hashidzume, H. Yamaguchi, Macroscopic self-assembly through molecular recognition. *Nature Chemistry* **3**, 34-37 (2011).
3. Q. Sun, R. Zhang, J. Qiu, R. Liu, W. Xu, On-Surface Synthesis of Carbon Nanostructures. *Advanced Materials* **30**, 1705630 (2018).
4. Z. Wang, K. Qian, M. A. Öner, P. S. Deimel, Y. Wang, S. Zhang, X. Zhang, V. Gupta, J. Li, H.-J. Gao, D. A. Duncan, J. V. Barth, X. Lin, F. Allegretti, S. Du, C.-A. Palma, Layer-by-Layer Epitaxy of Porphyrin−Ligand Fe(II)-Fe(III) Nanoarchitectures for Advanced Metal–Organic Framework Growth. *ACS Applied Nano Materials* **3**, 11752-11759 (2020).
5. L. Yan, O. J. Silveira, B. Alldritt, S. Kezilebieke, A. S. Foster, P. Liljeroth, Two-Dimensional Metal–Organic Framework on Superconducting $NbSe_2$. *ACS Nano* **15**, 17813-17819 (2021).
6. L. Yan, O. J. Silveira, B. Alldritt, O. Krejčí, A. S. Foster, P. Liljeroth, Synthesis and Local Probe Gating of a Monolayer Metal-Organic Framework. *Advanced Functional Materials* **31**, 2100519 (2021).
7. L. Hernández-López, I. Piquero-Zulaica, C. A. Downing, M. Piantek, J. Fujii, D. Serrate, J. E. Ortega, F. Bartolomé, J. Lobo-Checa, Searching for kagome multi-bands and edge states in a predicted organic topological insulator. *Nanoscale* **13**, 5216-5223 (2021).
8. B. Lowe, B. Field, J. Hellerstedt, J. Ceddia, H. L. Nourse, B. J. Powell, N. V. Medhekar, A. Schiffrin, Local gate control of Mott metal-insulator transition in a 2D metal-organic framework. *Nature Communications* **15**, 3559 (2024).
9. Z. F. Wang, Z. Liu, F. Liu, Organic topological insulators in organometallic lattices. *Nature Communications* **4**, 1471 (2013).
10. L. Z. Zhang, Z. F. Wang, B. Huang, B. Cui, Z. Wang, S. X. Du, H. J. Gao, F. Liu, Intrinsic Two-Dimensional Organic Topological Insulators in Metal–Dicyanoanthracene Lattices. *Nano Letters* **16**, 2072-2075 (2016).
11. B. Cui, X. Zheng, J. Wang, D. Liu, S. Xie, B. Huang, Realization of Lieb lattice in covalent-organic frameworks with tunable topology and magnetism. *Nature Communications* **11**, 66 (2020).
12. X. Ni, H. Li, F. Liu, J.-L. Brédas, Engineering of flat bands and Dirac bands in two-dimensional covalent organic frameworks (COFs): relationships among molecular orbital symmetry, lattice symmetry, and electronic-structure characteristics. *Materials Horizons* **9**, 88-98 (2022).
13. D. Kumar, J. Hellerstedt, B. Field, B. Lowe, Y. Yin, N. V. Medhekar, A. Schiffrin, Manifestation of Strongly Correlated Electrons in a 2D Kagome Metal–Organic Framework. *Advanced Functional Materials* **31**, 2106474 (2021).
14. X. Zhang, X. Li, Z. Cheng, A. Chen, P. Wang, X. Wang, X. Lei, Q. Bian, S. Li, B. Yuan, J. Gao, F.-S. Li, M. Pan, F. Liu, Large-scale 2D heterostructures from hydrogen-bonded organic frameworks and graphene with distinct Dirac and flat bands. *Nature Communications* **15**, 5934 (2024).

15. M. Garnica, D. Stradi, S. Barja, F. Calleja, C. Díaz, M. Alcamí, N. Martín, A. L. Vázquez de Parga, F. Martín, R. Miranda, Long-range magnetic order in a purely organic 2D layer adsorbed on epitaxial graphene. *Nature Physics* **9**, 368-374 (2013).
16. J. Liu, N. Lin, On-Surface-Assembled Single-Layer Metal-Organic Frameworks with Extended Conjugation. *ChemPlusChem* **88**, e202200359 (2023).
17. D. Baranowski, M. Thaler, D. Brandstetter, A. Windischbacher, I. Cojocariu, S. Mearini, V. Chesnyak, L. Schio, L. Floreano, C. Gutiérrez Bolaños, P. Puschnig, L. L. Patera, V. Feyer, C. M. Schneider, Emergence of Band Structure in a Two-Dimensional Metal–Organic Framework upon Hierarchical Self-Assembly. *ACS Nano* **18**, 19618-19627 (2024).
18. W.-C. Pan, C. Mützel, S. Haldar, H. Hohmann, S. Heinze, J. M. Farrell, R. Thomale, M. Bode, F. Würthner, J. Qi, Diboraperylene Diborinic Acid Self-Assembly on Ag(111)—Kagome Flat Band Localized States Imaged by Scanning Tunneling Microscopy and Spectroscopy. *Angewandte Chemie International Edition* **63**, e202400313 (2024).
19. X. Y. Zhu, Electronic structure and electron dynamics at molecule–metal interfaces: implications for molecule-based electronics. *Surface Science Reports* **56**, 1-83 (2004).
20. R. Temirov, S. Soubatch, A. Luican, F. S. Tautz, Free-electron-like dispersion in an organic monolayer film on a metal substrate. *Nature* **444**, 350-353 (2006).
21. T.-C. Tseng, C. Urban, Y. Wang, R. Otero, S. L. Tait, M. Alcamí, D. Écija, M. Trelka, J. M. Gallego, N. Lin, M. Konuma, U. Starke, A. Nefedov, A. Langner, C. Wöll, M. Á. Herranz, F. Martín, N. Martín, K. Kern, R. Miranda, Charge-transfer-induced structural rearrangements at both sides of organic/metal interfaces. *Nature Chemistry* **2**, 374-379 (2010).
22. Z. Li, B. Li, J. Yang, J. G. Hou, Single-molecule chemistry of metal phthalocyanine on noble metal surfaces. *Accounts of chemical research* **43**, 954-962 (2010).
23. A. Mugarza, R. Robles, C. Krull, R. Korytár, N. Lorente, P. Gambardella, Electronic and magnetic properties of molecule-metal interfaces: Transition-metal phthalocyanines adsorbed on Ag(100). *Physical Review B* **85**, 155437 (2012).
24. G. Koller, S. Berkebile, M. Oehzelt, P. Puschnig, C. Ambrosch-Draxl, F. P. Netzer, M. G. Ramsey, Intra- and Intermolecular Band Dispersion in an Organic Crystal. *Science* **317**, 351-355 (2007).
25. M. Wießner, J. Ziroff, F. Forster, M. Arita, K. Shimada, P. Puschnig, A. Schöll, F. Reinert, Substrate-mediated band-dispersion of adsorbate molecular states. *Nature Communications* **4**, 1514 (2013).
26. J. Hellerstedt, M. Castelli, A. Tadich, A. Grubišić-Čabo, D. Kumar, B. Lowe, S. Gicev, D. Potamianos, M. Schnitzenbaumer, P. Scigalla, S. Ghan, R. Kienberger, M. Usman, A. Schiffrin, Direct observation of narrow electronic energy band formation in 2D molecular self-assembly. *Nanoscale Advances* **4**, 3845-3854 (2022).
27. E. C. H. Wen, P. H. Jacobse, J. Jiang, Z. Wang, S. G. Louie, M. F. Crommie, F. R. Fischer, Fermi-Level Engineering of Nitrogen Core-Doped Armchair

Graphene Nanoribbons. *Journal of the American Chemical Society* **145**, 19338-19346 (2023).

28. M. Oehzelt, N. Koch, G. Heimel, Organic semiconductor density of states controls the energy level alignment at electrode interfaces. *Nature Communications* **5**, 4174 (2014).
29. M. Hollerer, D. Lüftner, P. Hurdax, T. Ules, S. Soubatch, F. S. Tautz, G. Koller, P. Puschnig, M. Sterrer, M. G. Ramsey, Charge Transfer and Orbital Level Alignment at Inorganic/Organic Interfaces: The Role of Dielectric Interlayers. *ACS Nano* **11**, 6252-6260 (2017).
30. Y. Chen, I. Tamblyn, S. Y. Quek, Energy Level Alignment at Hybridized Organic–Metal Interfaces: The Role of Many-Electron Effects. *The Journal of Physical Chemistry C* **121**, 13125-13134 (2017).
31. X.-Y. Liu, X.-Y. Xie, W.-H. Fang, G. Cui, Theoretical Insights into Interfacial Electron Transfer between Zinc Phthalocyanine and Molybdenum Disulfide. *The Journal of Physical Chemistry A* **122**, 9587-9596 (2018).
32. A. K. Geim, I. V. Grigorieva, Van der Waals heterostructures. *Nature* **499**, 419-425 (2013).
33. A. Kumar, K. Banerjee, A. S. Foster, P. Liljeroth, Two-Dimensional Band Structure in Honeycomb Metal–Organic Frameworks. *Nano Letters* **18**, 5596-5602 (2018).
34. C. Lyu, Y. Gao, K. Zhou, M. Hua, Z. Shi, P.-N. Liu, L. Huang, N. Lin, On-Surface Self-Assembly Kinetic Study of Cu-Hexaazatriphenylene 2D Conjugated Metal–Organic Frameworks on Coinage Metals and $MoS_2$ Substrates. *ACS Nano* **18**, 19793-19801 (2024).
35. P. Järvinen, S. K. Hämäläinen, K. Banerjee, P. Häkkinen, M. Ijäs, A. Harju, P. Liljeroth, Molecular Self-Assembly on Graphene on $SiO_2$ and h-BN Substrates. *Nano Letters* **13**, 3199-3204 (2013).
36. R. Barhoumi, A. Amokrane, S. Klyatskaya, M. Boero, M. Ruben, J.-P. Bucher, Screening the 4f-electron spin of $TbPc_2$ single-molecule magnets on metal substrates by ligand channeling. *Nanoscale* **11**, 21167-21179 (2019).
37. S. M. Obaidulla, M. R. Habib, Y. Khan, Y. Kong, T. Liang, M. Xu, $MoS_2$ and Perylene Derivative Based Type-II Heterostructure: Bandgap Engineering and Giant Photoluminescence Enhancement. *Advanced Materials Interfaces* **7**, 1901197 (2020).
38. Y. Guo, L. Wu, J. Deng, L. Zhou, W. Jiang, S. Lu, D. Huo, J. Ji, Y. Bai, X. Lin, S. Zhang, H. Xu, W. Ji, C. Zhang, Band alignment and interlayer hybridization in monolayer organic/$WSe_2$ heterojunction. *Nano Research* **15**, 1276-1281 (2022).
39. C. R. Lien-Medrano, F. P. Bonafé, C. Y. Yam, C.-A. Palma, C. G. Sánchez, T. Frauenheim, Fano Resonance and Incoherent Interlayer Excitons in Molecular van der Waals Heterostructures. *Nano Letters* **22**, 911-917 (2022).
40. S. Trishin, C. Lotze, N. Bogdanoff, F. von Oppen, K. J. Franke, Moire Tuning of Spin Excitations: Individual Fe Atoms on $MoS_2$/Au(111). *Physical Review Letters* **127**, 236801 (2021).
41. J. Lobo-Checa, L. Hernández-López, M. M. Otrokov, I. Piquero-Zulaica, A. E. Candia, P. Gargiani, D. Serrate, F. Delgado, M. Valvidares, J. Cerdá, A. Arnau, F.

Bartolomé, Ferromagnetism on an atom-thick & extended 2D metal-organic coordination network. *Nature Communications* **15**, 1858 (2024).

42. G. Reecht, N. Krane, C. Lotze, L. Zhang, A. L. Briseno, K. J. Franke, Vibrational Excitation Mechanism in Tunneling Spectroscopy beyond the Franck-Condon Model. *Physical Review Letters* **124**, 116804 (2020).

43. C.-A. Palma, S. Joshi, T. Hoh, D. Ecija, J. V. Barth, W. Auwärter, Two-Level Spatial Modulation of Vibronic Conductance in Conjugated Oligophenylenes on Boron Nitride. *Nano Letters* **15**, 2242-2248 (2015).

44. Y. J. Zheng, Y. L. Huang, Y. Chen, W. Zhao, G. Eda, C. D. Spataru, W. Zhang, Y.-H. Chang, L.-J. Li, D. Chi, Heterointerface screening effects between organic monolayers and monolayer transition metal dichalcogenides. *ACS nano* **10**, 2476-2484 (2016).

45. B. Han, P. Samorì, Engineering the Interfacing of Molecules with 2D Transition Metal Dichalcogenides: Enhanced Multifunctional Electronics. *Accounts of Chemical Research*, (2024).

46. W. Jiang, H. Huang, F. Liu, A Lieb-like lattice in a covalent-organic framework and its Stoner ferromagnetism. *Nature Communications* **10**, 2207 (2019).

47. S. Gottardi, K. Müller, J. C. Moreno-López, H. Yildirim, U. Meinhardt, M. Kivala, A. Kara, M. Stöhr, Cyano-Functionalized Triarylamines on Au (111): Competing Intermolecular versus Molecule/Substrate Interactions. *Advanced Materials Interfaces* **1**, 1300025 (2014).

48. K. Müller, J. C. Moreno-López, S. Gottardi, U. Meinhardt, H. Yildirim, A. Kara, M. Kivala, M. Stöhr, Cyano-Functionalized Triarylamines on Coinage Metal Surfaces: Interplay of Intermolecular and Molecule–Substrate Interactions. *Chemistry–A European Journal* **22**, 581-589 (2016).

49. Y. Okuno, T. Yokoyama, S. Yokoyama, T. Kamikado, S. Mashiko, Theoretical Study of Benzonitrile Clusters in the Gas Phase and Their Adsorption onto a Au(111) Surface. *Journal of the American Chemical Society* **124**, 7218-7225 (2002).

50. M. Bieri, S. Blankenburg, M. Kivala, C. A. Pignedoli, P. Ruffieux, K. Müllen, R. Fasel, Surface-supported 2D heterotriangulene polymers. *Chemical Communications* **47**, 10239-10241 (2011).

51. M. Raya-Moreno, C. Cazorla, E. Canadell, R. Rurali, Phonon transport manipulation in TiSe$_2$ via reversible charge density wave melting. *npj 2D Materials and Applications* **8**, 64 (2024).

52. M. Spera, A. Scarfato, A. Pasztor, E. Giannini, D. R. Bowler, C. Renner, Insight into the charge density wave gap from contrast inversion in topographic STM images. *Physical Review Letters* **125**, 267603 (2020).

53. C. Steiner, J. Gebhardt, M. Ammon, Z. Yang, A. Heidenreich, N. Hammer, A. Görling, M. Kivala, S. Maier, Hierarchical on-surface synthesis and electronic structure of carbonyl-functionalized one- and two-dimensional covalent nanoarchitectures. *Nature Communications* **8**, 14765 (2017).

54. C. Steiner, Z. Yang, B. D. Gliemann, U. Meinhardt, M. Gurrath, M. Ammon, B. Meyer, M. Kivala, S. Maier, Binary supramolecular networks of bridged triphenylamines with different substituents and identical scaffolds. *Chemical Communications* **54**, 11554-11557 (2018).

55. S. Maier, "Alterations in the Electronic Structure Upon Hierarchical Growth of 2D Networks" in *Encyclopedia of Interfacial Chemistry,* K. Wandelt, Ed. (Elsevier, Oxford, 2018), pp. 195-203.
56. W. Liu, A. Tkatchenko, M. Scheffler, Modeling Adsorption and Reactions of Organic Molecules at Metal Surfaces. *Accounts of Chemical Research* **47**, 3369-3377 (2014).
57. M. Palummo, A. N. D'Auria, J. C. Grossman, G. Cicero, Tailoring the optical properties of $MoS_2$ and $WS_2$ single layers via organic functionalization. *Journal of Physics: Condensed Matter* **31**, 235701 (2019).
58. O. Adeniran, Z.-F. Liu, Quasiparticle electronic structure of phthalocyanine: TMD interfaces from first-principles GW. *The Journal of Chemical Physics* **155**, (2021).
59. I. Gonzalez Oliva, F. Caruso, P. Pavone, C. Draxl, Hybrid excitations at the interface between a $MoS_2$ monolayer and organic molecules: A first-principles study. *Physical Review Materials* **6**, 054004 (2022).
60. N. Koch, Energy levels at interfaces between metals and conjugated organic molecules. *Journal of Physics: Condensed Matter* **20**, 184008 (2008).
61. L. Camilli, C. Hogan, D. Romito, L. Persichetti, A. Caporale, M. Palummo, M. Di Giovannantonio, D. Bonifazi, On-Surface Molecular Recognition Driven by Chalcogen Bonding. *JACS Au*, (2024).
62. S. Wickenburg, J. Lu, J. Lischner, H.-Z. Tsai, A. A. Omrani, A. Riss, C. Karrasch, A. Bradley, H. S. Jung, R. Khajeh, D. Wong, K. Watanabe, T. Taniguchi, A. Zettl, A. H. C. Neto, S. G. Louie, M. F. Crommie, Tuning charge and correlation effects for a single molecule on a graphene device. *Nature Communications* **7**, 13553 (2016).
63. M. Wiesenmayer, S. Hilgenfeldt, S. Mathias, F. Steeb, T. Rohwer, M. Bauer, Spectroscopy and population decay of a van der Waals gap state in layered $TiSe_2$. *Physical Review B* **82**, 035422 (2010).
64. M. D. Watson, O. J. Clark, F. Mazzola, I. Marković, V. Sunko, T. K. Kim, K. Rossnagel, P. D. C. King, Orbital- and $k_z$-Selective Hybridization of Se $4p$ and Ti $3d$ States in the Charge Density Wave Phase of $TiSe_2$. *Physical Review Letters* **122**, 076404 (2019).
65. D. Nečas, P. Klapetek, Gwyddion: an open-source software for SPM data analysis. *Open Physics* **10**, 181-188 (2012).
66. V. Blum, R. Gehrke, F. Hanke, P. Havu, V. Havu, X. Ren, K. Reuter, M. Scheffler, Ab initio molecular simulations with numeric atom-centered orbitals. *Computer Physics Communications* **180**, 2175-2196 (2009).
67. P. Hohenberg, W. Kohn, Inhomogeneous Electron Gas. *Physical Review* **136**, B864-B871 (1964).
68. W. Kohn, L. J. Sham, Self-Consistent Equations Including Exchange and Correlation Effects. *Physical Review* **140**, A1133-A1138 (1965).
69. A. Gulans, S. Kontur, C. Meisenbichler, D. Nabok, P. Pavone, S. Rigamonti, S. Sagmeister, U. Werner, C. Draxl, exciting: a full-potential all-electron package implementing density-functional theory and many-body perturbation theory. *Journal of Physics: Condensed Matter* **26**, 363202 (2014).

70. J. P. Perdew, K. Burke, M. Ernzerhof, Generalized Gradient Approximation Made Simple. *Physical Review Letters* **77**, 3865-3868 (1996).
71. A. Tkatchenko, M. Scheffler, Accurate Molecular Van Der Waals Interactions from Ground-State Electron Density and Free-Atom Reference Data. *Physical Review Letters* **102**, 073005 (2009).
72. C. Draxl, M. Scheffler, The NOMAD laboratory: from data sharing to artificial intelligence. *Journal of Physics: Materials* **2**, 036001 (2019).

Supplementary Materials for

# "Zero-energy band observation in an interfacial chalcogen-organic network"

Yichen Jin *et al.*

*Corresponding author: Carlos-Andres Palma, Email: palma@physik.hu-berlin.de

**This PDF file includes:**

    Supplementary Text
    Figs. S1 to S6
    References (1 to 16)

**Supplementary Text**

    In particular, molecule-based networks fabricated through covalent organic, metal-organic and hydrogen-bonded strategies have led to the observation of Dirac bands and flat bands which have been predicted to lead to topological states. More information about flat bands, Dirac bands and topological bands is given in the following references: (*1-14*).

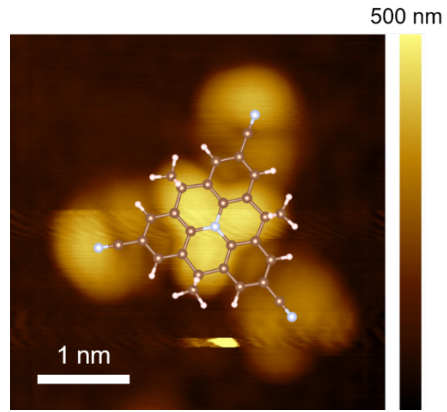

**Fig. S1. Topography of single triarylamine on TiSe$_2$**

High-resolution STM image of single triarylamine on top of TiSe$_2$. (set-point: 2 pA, U$_{sample\_bias}$ = 100 mV and temperature: 5.3 K)

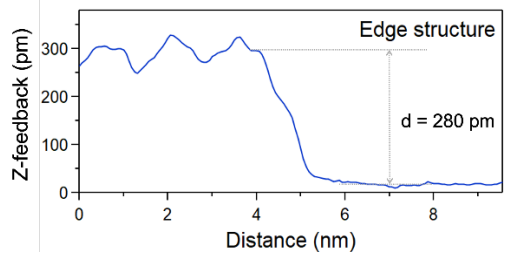

**Fig. S2. Height of triarylamine island**
Height profile corresponding to molecule edge line in Fig. 2A.

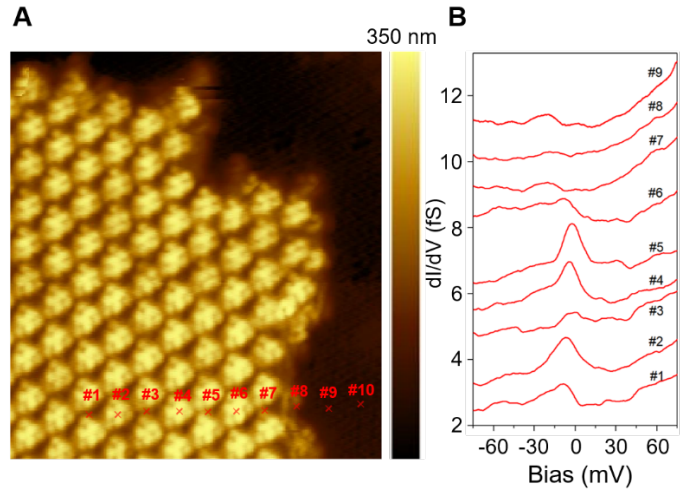

**Fig. S3. Additional STS measurements at different positions.**
(**A**) The STM scanning area with marks demonstrates the STS measurement positions.
(**B**) More STS data on triarylamine in the heterostructure at the marked position (Setpoint: 5 pA, $U_{sample\_bias}$ = -300 mV, $U_{ac}$ = 2 mV, 973 Hz and T = 5.1 K).

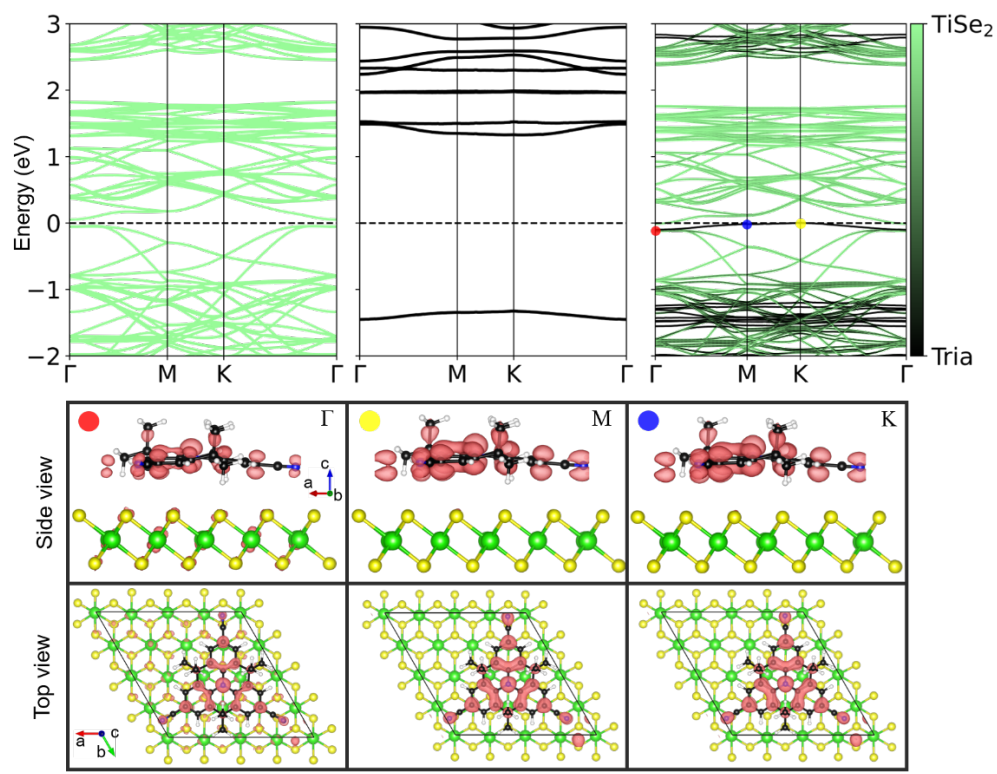

**Fig. S4. The respective electronic structures of triarylamine and TiSe$_2$**
Projected band structures for pure TiSe$_2$ (top left), triarylamine network (top middle), and triarylamine on TiSe$_2$ (top right) systems. The reference Fermi level is automatically selected as an intermediate position between LOMOs and HOMO in the triarylamine network calculation. Top and side view of projected KS wave functions (red) of the triarylamine (HOMO) flat band at the Γ, K, and M points with isosurface: $0.001 e·Å^{-3}$.

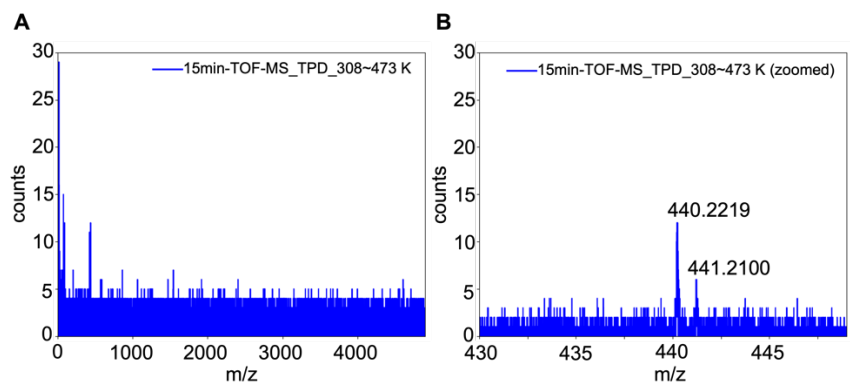

**Fig. S5. Triarylamine purity characterization**

Overview (**A**) and zoomed-in (**B**) TOF-MS of triarylamine during TPD heating from 308 K to 473 K for 15 min. m/z = 440.2219, 441.2100 correspond to triarylamine (simulation m/z = 440.2001, 441.2035).

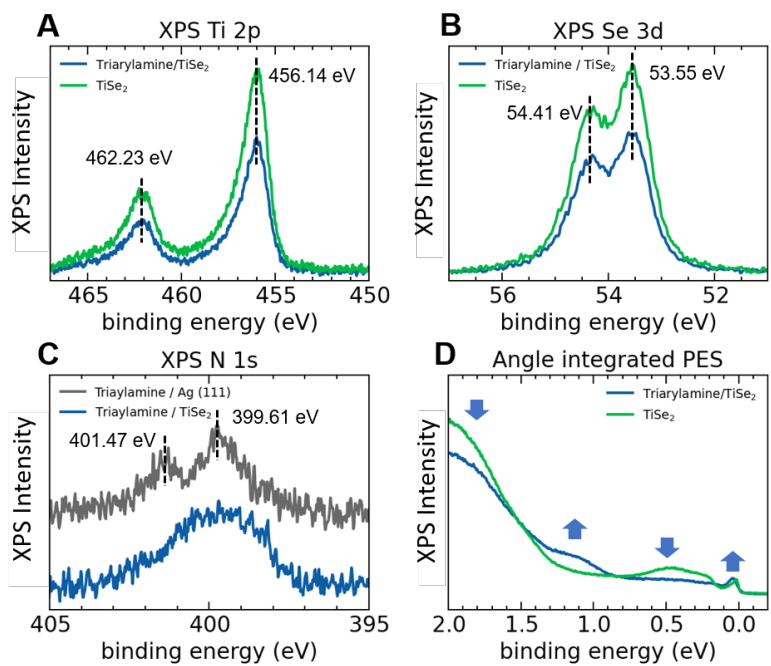

**Fig. S6. Cole level electronic structure.**
XPS spectra (Shirley type backgrounds are subtracted) collected of submonolayer triarylamine on TiSe$_2$ at 15 K: **(A)** Ti 2$p$, **(B)** Se 3$d$, and **(C)** N 1$s$ with a reference sign on Ag(111) surface of $h\nu$ = 1253.6 eV. Angle integrated PES spectra collected of submonolayer triarylamine on TiSe$_2$ at 15 K: **(D)** valence band of He-I: 21.2 eV.

In Fig. S5A and S5B, the Ti 2$p$ and Se 3$d$ peaks show only a slight decrease in Ti 2$p$ and Se 3$d$ intensity without significant chemical shift, indicating very weak orbital hybridization. The reduction in XPS intensity is attributed to scattering by the molecular. Further, in Fig. S5C, the N 1$s$ peak confirms the presence of the molecule. By performing peak fitting, peaks located at binding energy (E$_b$) ~ 400 eV are resolved into two distinct N chemical environments in a ratio of 2.98:1. As the single triarylamine STM data and its stick model (Fig. S1), the nitrogen atoms have center and edge position. The center nitrogen atoms (N–2) are bonded to three adjacent carbon atoms, forming three C–N single bonds (CN$_1$), while in the three CN$_3$ moieties, the nitrogen (N–1, N–3, and N–4) connected via a C≡N triple. The ratio of CN$_3$ to CN$_3$ moieties aligns with the fitting results. Although exhibiting slight deviations from the XPS peak shape of triarylamine on Ag(111), the structure and properties of Ag(111) have been extensively discussed in Ref(*15*), where the edge CN$_3$ faces the neighboring benzene ring. The broadened N 1$s$ peaks on TiSe$_2$, compared to those on Ag(111), are attributed to the crystallization structures of triarylamine on different substrates, the distance between neighboring edge CN$_3$ measures only 2.24 Å (ref(*16*): N-N bond in N$_2$, 1.45 Å). Nonetheless, the broadened N 1$s$ peaks suggest some possibility of connection or hybridization among the CN$_3$ moieties. All angle integrated PES characteristics were evaluated (Fig. S5D) with submonolayer coverage. The signal at E$_b$ ~ 0.5 eV is attenuated while the spectra at E$_b$ ~ 0.1 and 1.2 eV are enhanced after the deposition.

# References

1. P. Zhu, V. Meunier, Electronic properties of two-dimensional covalent organic frameworks. *The Journal of Chemical Physics* **137**, (2012).
2. Z. Liu, Z.-F. Wang, J.-W. Mei, Y.-S. Wu, F. Liu, Flat Chern Band in a Two-Dimensional Organometallic Framework. *Physical Review Letters* **110**, 106804 (2013).
3. Z. F. Wang, Z. Liu, F. Liu, Organic topological insulators in organometallic lattices. *Nature Communications* **4**, 1471 (2013).
4. Z. F. Wang, N. Su, F. Liu, Prediction of a Two-Dimensional Organic Topological Insulator. *Nano Letters* **13**, 2842-2845 (2013).
5. Exotic electronic states in the world of flat bands: From theory to material. *Chinese Physics B* **23**, 077308 (2014).
6. C. Barreteau, F. Ducastelle, T. Mallah, A bird's eye view on the flat and conic band world of the honeycomb and Kagome lattices: towards an understanding of 2D metal-organic frameworks electronic structure. *Journal of Physics: Condensed Matter* **29**, 465302 (2017).
7. W. Jiang, H. Huang, F. Liu, A Lieb-like lattice in a covalent-organic framework and its Stoner ferromagnetism. *Nature Communications* **10**, 2207 (2019).
8. Y. Jing, T. Heine, Two-Dimensional Kagome Lattices Made of Hetero Triangulenes Are Dirac Semimetals or Single-Band Semiconductors. *Journal of the American Chemical Society* **141**, 743-747 (2019).
9. S. Thomas, H. Li, C. Zhong, M. Matsumoto, W. R. Dichtel, J.-L. Bredas, Electronic Structure of Two-Dimensional π-Conjugated Covalent Organic Frameworks. *Chemistry of Materials* **31**, 3051-3065 (2019).
10. Y. Gao, Y.-Y. Zhang, J.-T. Sun, L. Zhang, S. Zhang, S. Du, Quantum anomalous Hall effect in two-dimensional Cu-dicyanobenzene coloring-triangle lattice. *Nano Research* **13**, 1571-1575 (2020).
11. X. Ni, Y. Zhou, G. Sethi, F. Liu, π-Orbital Yin–Yang Kagome bands in anilato-based metal–organic frameworks. *Physical Chemistry Chemical Physics* **22**, 25827-25832 (2020).
12. D. J. Rizzo, Q. Dai, C. Bronner, G. Veber, B. J. Smith, M. Matsumoto, S. Thomas, G. D. Nguyen, P. R. Forrester, W. Zhao, J. H. Jørgensen, W. R. Dichtel, F. R. Fischer, H. Li, J.-L. Bredas, M. F. Crommie, Revealing the Local Electronic Structure of a Single-Layer Covalent Organic Framework through Electronic Decoupling. *Nano Letters* **20**, 963-970 (2020).
13. Y. Zhou, G. Sethi, C. Zhang, X. Ni, F. Liu, Giant intrinsic circular dichroism of enantiomorphic flat Chern bands and flatband devices. *Physical Review B* **102**, 125115 (2020).
14. W. Jiang, X. Ni, F. Liu, Exotic Topological Bands and Quantum States in Metal–Organic and Covalent–Organic Frameworks. *Accounts of Chemical Research* **54**, 416-426 (2021).
15. K. Müller, J. C. Moreno-López, S. Gottardi, U. Meinhardt, H. Yildirim, A. Kara, M. Kivala, M. Stöhr, Cyano-Functionalized Triarylamines on Coinage Metal Surfaces: Interplay of Intermolecular and Molecule–Substrate Interactions. *Chemistry–A European Journal* **22**, 581-589 (2016).

16. C. Chen, S.-F. Shyu, Theoretical study of single-bonded nitrogen cluster-type molecules. *International Journal of Quantum Chemistry* **73**, 349-356 (1999).